\documentstyle[12pt]{article}

\begin{document}

\title{Computing Casimir invariants from Pffafian systems}
\author{T. W. Yudichak$^{\;(1)}$ \and Benito Hern\'{a}ndez--Bermejo$^{\;(2)}$ \and %
P. J. Morrison$^{\;(1,*)}$}
\date{}
\maketitle

\begin{abstract}
We describe a method for computing Casimir invariants that is applicable to
both finite and infinite-dimensional Poisson brackets. We apply the method
to various finite and infinite-dimensional examples, including a Poisson
bracket embodying both finite and infinite-dimensional structure.
\end{abstract}

\setlength{\baselineskip}{18pt} 

\noindent {\em $^1$ Department of Physics and Institute for Fusion Studies,
The University of Texas at Austin, Austin, TX 78712, USA}

\noindent {\em $^2$Departamento de F\'{\i}sica Matem\'{a}tica y Fluidos,
Universidad Nacional de Educaci\'{o}n a Distancia. Senda del Rey S/N, 28040
Madrid, Spain. \newline
E-mail: bhernand@apphys.uned.es}

\mbox{}

\mbox{}

\mbox{}


\bigskip


\mbox{}

\noindent $^*$ To whom all correspondence should be addressed.

\pagebreak

\section{Introduction}

\label{S-INTRO}

Noncanonical Hamiltonian, or Poisson, systems have been studied extensively
by mathematical physicists, and with good reason. They arise in fields as
diverse as dynamical systems theory \cite{gyn1,pla1}, fluid dynamics and
plasma physics \cite{flpl,olv3}, and condensed matter physics \cite{cmph}.
(For more references see \cite{mara,mrmp}.)

One of the key features of Hamiltonian systems is the existence of a Poisson
bracket. This object finds many applications, including computing
perturbative solutions \cite{pert}, determining stability of steady states 
\cite{stss}, and studying integrable systems \cite{pla1,intg,kosc}.

The ubiquity of the Poisson bracket has led to study of it in its own right.
In particular, a good deal of work has gone into exploring its geometric
theory \cite{wei1,vais} (also see references on the related subject of
symplectic geometry \cite{syge}). In this letter, we consider only one facet
of this rich geometry, that related to Casimir invariants. These are
functions whose Poisson bracket with any other function vanishes. Their
existence for any finite-dimensional degenerate Poisson bracket follows from
the Frobenius theorem of differential geometry \cite{frob}.

The Casimir invariants of a given Poisson bracket are important because they
are conserved quantities in any Hamiltonian system that uses that bracket.
As such, they play a vital part in reducing the order \cite{red}, or even
integrating some systems \cite{mmof}. In addition, they are central to both
the Energy-Casimir method of determining stability \cite{stss}, and the
Semenov-Tian-Shansky scheme of constructing integrable systems \cite{kosc}.
In all these applications, the actual computation of Casimir invariants is
essential.

We present a method for computing Casimir invariants. This method amounts to
integrating the null covectors of the second rank tensor (the cosymplectic
form) defined by the Poisson bracket. This is a natural strategy: many of
the Casimir invariants in \cite{mrmp} were calculated in this fashion, a
fact not noted there. This method {\it has} been presented explicitly, for
infinite-dimensional brackets in \cite{kuro} and finite-dimensional brackets
in \cite{byv2}. Here, we show that these papers both treat special cases of
a single method by presenting the method in a geometrical framework that
includes both cases (Section \ref{S-CAS}). This presentation has a partly
pedagogical function: we present a transparent example of differential
geometry applied to Hamiltonian dynamics. But also, it offers a clear
framework for arriving at a novel result: the form of Casimir invariants for
Poisson brackets with a nested Lie-Poisson structure, involving both finite
and infinite-dimensional brackets. The method is used to compute these
Casimir invariants (and those of several other Poisson brackets) in Sections 
\ref{S-FINITE} and \ref{S-INFINITE}.

\section{Casimir invariants and Pfaffian systems}

\label{S-CAS}


The Poisson bracket is a mathematical structure common to all Hamiltonian
systems. If $z^{\alpha}$ denotes a coordinate of a point in a phase space M,
the equations of motion generated by a Hamiltonian $H \colon {\rm M}
\rightarrow I \!\! R$ are $\dot z^{\alpha} = \{z^{\alpha}, H \}$ where $\{
\cdot , \cdot \} \colon C^{\infty} ({\rm M}) \times C^{\infty} ({\rm M})
\rightarrow C^{\infty} ({\rm M}).$

The bracket that appears in the equation of motion is a Poisson bracket if
it is antisymmetric, bilinear, and it satisfies two more condtions. These
are the Leibnitz rule, $\{ f,gh \} = g \{ f,h \} + \{ f,g \} h$ and the
Jacobi identity,

\begin{equation}  \label{JACOBI}
\{ f, \{ g,h \} \} + \{ g, \{ h,f \} \} + \{ h, \{ f,g \} \} = 0.
\end{equation}

The most familiar example of a bracket satisfying these properties is given
by the so-called canonical bracket:

\begin{equation}  \label{CANBRAC}
\{ f, g \} = \sum_{i=1}^n {\frac{\partial f }{\partial q^i}} {\frac{\partial
g }{\partial p_i}} -{\frac{\partial g }{\partial q^i}} {\frac{\partial f }{%
\partial p_i}}.
\end{equation}
But there are the other, noncanonical brackets that obey the above
conditions; some are given in the examples. However, such noncanonical
brackets, even when they are degenerate, can be locally represented in the
form (\ref{CANBRAC}) (where $n$ is equal to half the rank of the bracket),
as shown by the Darboux theorem (see also \cite{wei1}). In this sense, they
are true generalizations of the more familiar canonical brackets.

By virtue of the above properties, a Poisson bracket defines a second rank
tensor called the cosymplectic form, ${\rm J}\colon {\rm T^*M} \rightarrow 
{\rm TM}$, in the following way

\begin{equation}  \label{JDEF}
\langle {\rm d}f(z), {\rm J}(z) {\rm d}g(z) \rangle \colon = \{ f, g \} (z)
\end{equation}
where the angular brackets denote the pairing between the cotangent and
tangent spaces at the point $z \in {\rm M}.$ In a given set of coordinates $%
z^{\alpha },$ for a finite-dimensional system, equation (\ref{JDEF}) has the
form

\begin{equation}  \label{JDEF2}
{\frac{\partial f }{\partial z^{\alpha }}} {\rm J}^{\alpha \beta} {\frac{%
\partial g }{\partial z^{\beta }}} \colon = \{ f,g \} \, .
\end{equation}
(We use the summation convention.) The vector field given by Jd$g$ is called
the Hamiltonian vector field generated by $g.$


Casimir invariants are functions that have a zero Poisson bracket with any
other function. In other words, given a Poisson bracket, if, for every $g,$
a function $C$ satisfies $\{ g,C \} = 0,$ then $C$ is a Casimir invariant.

This definition is the starting point of any method to compute Casimir
invariants. In terms of the tensor J, it becomes $\langle {\rm d}g, {\rm J}%
{\rm d}C \rangle = 0.$ Since $g$ is arbitrary, solving this equation amounts
to finding the functions $C$ from the condition that the vector field ${\rm J%
}{\rm d}C$ vanishes. In the case of a finite-dimensional system, this
computation requires solving a set of coupled partial differential equations
for the function $C$: ${\rm J}^{ij} \partial C / \partial z^j =0$. However,
a more efficient method for computing Casimir invariants follows from an
alternate geometric interpretation of $\{ g,C \}=0.$ Making use of the
antisymmetry of the Poisson bracket, we rewrite $\langle {\rm d}g,{\rm J}%
{\rm d}C \rangle = 0$ as $\langle {\rm d}C,{\rm J}{\rm d}g \rangle = 0.$
Since this equation must hold for arbitrary $g$, we can interpret it as
saying that the differential one-form given by ${\rm d}C$ is annulled by
every vector field in the image of J.

Geometrically, this means that the vector field Jd$g$ must lie everywhere
tangent to the level set of the function $C$. In this case, any integral
curve of this (or any other) Hamiltonian vector field must lie within a
single level set of $C$, which is a way of saying that $C$ is a constant of
motion.

These considerations help us to compute Casimir invariants in the following
way. The information needed to construct the level sets of Casimir
invariants is contained in the differential one-forms that are annuled by
the vectors tangent to these sets. Obviously, if we explicitly knew a
Casimir invariant $C$, $\langle {\rm d}C,{\rm J}{\rm d}g \rangle = 0$ tells
us that d$C$ is such a one-form. However, we do not need to know the Casimir
invariants to find differential one-forms that are annulled by these
vectors. In fact, from $\langle {\rm d}g,{\rm J}{\rm d}C \rangle = \langle 
{\rm d}C,{\rm J}{\rm d}g \rangle = 0,$ it is clear that any element in the
kernel of J will do. And so, we can get the information we need to determine
all of the Casimir invariants by finding all the linearly independent null
covectors of J.


The null covectors of J, call them $\gamma ^{(i)}$ (suppose there are $n$),
are differential one-forms. They are not necessarily exact: that is, it is
not necessarily true that there exists an $F$ such that $\gamma ^{(i)} = 
{\rm d}F.$

But once we have the $\gamma ^{(i)}$, we can find the Casimir invariants
from the condition that the $\gamma ^{(i)}$ vanish when paired with any
vector tangent to a level set of a Casimir invariant: $\gamma ^{(i)} = 0$
where $i$ varies from $1$ to $n.$ A system of equations, like this one,
given by setting a collection of one-forms to zero, is called a Pfaffian
system. The level sets of the functions on which the restrictions of these
one-forms vanish are called the integral manifolds of the system.

Finding the integral manifolds is conceptually (but not always
computationally) simple. If we were able to linearly combine the $\gamma
^{(i)}$ into $n$ independent exact differentials, we would obtain a system
of equations equivalent to $\gamma ^{(i)} = 0$. This system would have the
form ${\rm d}F^{(j)} = 0$ where $j$ varies from $1$ to $n.$ The solution to
this system of equations is obviously given by $F^{(j)} = {\rm constant}.$
In other words, the integral manifolds are just the level sets of the $%
F^{(j)}.$ By the chain rule, we can then write the general solution of $%
\gamma ^{(i)} = 0$ in terms of an arbitrary function $k = k(F^{(1)},\cdots
,F^{(n)}).$

It is not always possible to transform a Pfaffian system in this way. The
conditions for when it is possible are given by the Frobenius theorem. In 
\cite{frob} it was shown that the possibility of doing this with the set of
null covectors of a finite-dimensional J follows from equation (\ref{JACOBI}%
). Later, in \cite{kuro}, an argument was given that extends this proof to
continuum systems. However, there are some infinite-dimensional spaces in
which the Frobenius theorem does not hold (see \cite{krmi}), so this should
be seen as a formal argument.

Examples of applying the above process are given in the next two sections.
In the finite-dimensional case, the greatest complication is finding
integrating factors to make the resulting equations exact differential
equations (see Section \ref{S-FINITE}). The infinite-dimensional case is
typically more complicated. Finding components of the null covectors
requires solving partial differential equations, and it is not always
obvious how to integrate the Pfaffian system.

\section{Casimir invariants of finite-dimensional brackets}

\label{S-FINITE} 

For our first example, we will find the Casimir invariant of a familiar
Hamiltonian system: the free rigid body. The components of J are given by: $%
{\bf {\rm J}}^{ij} = \epsilon _k^{ij} z^k$ where $i,j = 1 \ldots 3,$ and $%
\epsilon$ denotes the Levi-Civita symbol, as usual. In fact, this bracket is
the Lie-Poisson bracket associated to the Lie algebra so(3) (see Section \ref
{S-INFINITE}). Note that rank(J) $= \; 2$ everywhere except at the origin.
The null covector of J is immediately found to be $\gamma = z^1{\rm d}z^1 +
z^2\mbox{d}z^2 + z^3\mbox{d}z^3.$ And the Pfaffian system $\gamma = 0$ is
immediately solved for the well-known Casimir invariant $C=
(z^1)^2+(z^2)^2+(z^3)^2 = \mbox{constant}.$

As a second example, we consider the brackets of arbitrary dimension $d$
introduced by Plank \cite{pla1} for Lotka-Volterra equations (see also \cite
{byv3} for an application of such brackets to more general systems). In this
case, the components of J are given by (not summed): 
\begin{equation}  \label{ND.1}
{\rm J}^{ij} = c_{(ij)} z^i z^j \quad \quad i,j = 1 \ldots d
\end{equation}
where $c_{(ij)} \in I \! \! R$ and $c_{(ij)} = - c_{(ji)}$ for all $i,j$. It
can be shown \cite{pla1} that the skew-symmetry of the $c_{ij}$ is enough to
ensure that J in (\ref{ND.1}) indeed defines a cosymplectic form. It is
simple to prove that every null covector of the matrix J in (\ref{ND.1}) is
of the form:

\begin{equation}  \label{ND.2}
\gamma = a_i \frac{\mbox{d}z^i}{z^i} \, , \:\;\:\;\:\; a_i \in I \!\! R \, .
\end{equation}

Every Pfaffian equation $\gamma =0$, with $\gamma$ given by expression (\ref
{ND.2}), can be integrated to give one Casimir invariant:

\begin{equation}  \label{ND.3}
C = a_i \log z^i = \mbox{\rm constant} \, .
\end{equation}
Therefore, all Casimir invariants of (\ref{ND.1}) can be easily found in
this way.

An example that involves solving a less trivial Pfaffian system is the
Lie-Poisson bracket associated with the Lie algebra su(3) (see Section \ref
{S-INFINITE}). This arises in the study of finite-mode analogs of
two-dimensional hydrodynamics \cite{zeit}. With respect to ($i$ times the)
Gell-Mann basis of su(3) \cite{gema}, the components of J are given by

\begin{equation}  \label{SU3-COSYM}
\left( 
\begin{array}{cccccccc}
0 & 2 z^3 & -2 z^2 & z^7 & -z^6 & z^5 & -z^4 & 0 \\ 
-2 z^3 & 0 & 2 z^1 & z^6 & z^7 & -z^4 & -z^5 & 0 \\ 
2 z^2 & -2 z^1 & 0 & z^5 & -z^4 & -z^7 & z^6 & 0 \\ 
-z^7 & -z^6 & -z^5 & 0 & z^3 + \sqrt{3} z^8 & z^2 & z^1 & -\sqrt{3} z^5 \\ 
z^6 & -z^7 & z^4 & -z^3 - \sqrt{3} z^8 & 0 & -z^1 & z^2 & \sqrt{3} z^4 \\ 
-z^5 & z^4 & z^7 & -z^2 & z^1 & 0 & -z^3 + \sqrt{3} z^8 & -\sqrt{3} z^7 \\ 
z^4 & z^5 & -z^6 & -z^1 & -z^2 & z^3 - \sqrt{3} z^8 & 0 & \sqrt{3} z^6 \\ 
0 & 0 & 0 & \sqrt{3} z^5 & -\sqrt{3} z^4 & \sqrt{3} z^7 & -\sqrt{3} z^6 & 0
\end{array}
\right) \, .
\end{equation}
The null vectors (and resulting Pfaffian system) of this J were found using 
{\it Mathematica} to be

\begin{eqnarray}
\gamma ^{(1)} & = & \Delta ^{-1} [2\sqrt{3}(z^1 z^2 z^4 - (z^1)^2 z^5 + z^1
z^3 z^7 + z^4 z^6 z^7 + z^5 (z^7)^2 + \sqrt{3} z^1 z^7 z^8){\rm d}z^1 
\nonumber \\
& + & 2\sqrt{3}((z^2)^2 z^4 - z^1 z^2 z^5 + z^2 z^3 z^7 + z^5 z^6 z^7 - z^4
(z^7)^2 + \sqrt{3} z^2 z^7 z^8){\rm d}z^2  \nonumber \\
& + & \sqrt{3}(2z^2 z^3 z^4 - 2 z^1 z^3 z^5 + 2 (z^3)^2 z^7 + (z^4)^2 z^7 +
(z^5)^2 z^7 - (z^6)^2 z^7  \nonumber \\
&& - (z^7)^3 + 2 \sqrt{3} z^3 z^7 z^8){\rm d}z^3  \nonumber \\
& + & 2\sqrt{3}(z^2 (z^4)^2 - z^1 z^4 z^5 + 2 z^3 z^4 z^7 + z^1 z^6 z^7 -
z^2 (z^7)^2){\rm d}z^4  \label{SU3-1} \\
& + & 2\sqrt{3}(z^2 z^4 z^5 - z^1 (z^5)^2 + 2 z^3 z^5 z^7 + z^2 z^6 z^7 +
z^1 (z^7)^2){\rm d}z^5  \nonumber \\
& + & 2\sqrt{3}(z^2 z^4 z^6 - z^1 z^5 z^6 + z^1 z^4 z^7 + z^2 z^5 z^7){\rm d}%
z^6  \nonumber \\
& + & (2(z^1)^2 z^7 + 2(z^2)^2 z^7 + 2(z^3)^2 z^7 - (z^4)^2 z^7 - (z^5)^2
z^7 - (z^6)^2 z^7  \nonumber \\
&& - (z^7)^3 + 2\sqrt{3} (z^2 z^4 - z^1 z^5 + z^3 z^7) z^8){\rm d}z^8]=0 \, ,
\nonumber
\end{eqnarray}
and 
\begin{eqnarray}
\gamma ^{(2)} & = & \Delta ^{-1} [(2(z^1)^3 + 2z^1 (z^2)^2 + 2 z^1 (z^3)^2 -
z^1 (z^4)^2 - z^1 (z^5)^2 - z^1 (z^6)^2 - z^1 (z^7)^2  \nonumber \\
&& - 2\sqrt{3} z^4 z^6 z^8 - 2\sqrt{3} z^5 z^7 z^8 - 6 z^1 (z^8)^2){\rm d}z^1
\nonumber \\
& + & (2(z^1)^2 z^2 + 2 (z^2)^3 + 2 z^2 (z^3)^2 - z^2 (z^4)^2 - z^2 (z^5)^2
- z^2 (z^6)^2 - z^2 (z^7)^2  \nonumber \\
&& - 2\sqrt{3} z^5 z^6 z^8 + 2\sqrt{3} z^4 z^7 z^8 - 6 z^2 (z^8)^2){\rm d}z^2
\nonumber \\
& + & (2(z^1)^2 z^3 + 2 (z^2)^2 z^3 + 2 (z^3)^3 - z^3 (z^4)^2 - z^3 (z^5)^2
- z^3 (z^6)^2 - z^3 (z^7)^2  \nonumber \\
&& - \sqrt{3} (z^4)^2 z^8 - \sqrt{3} (z^5)^2 z^8 + \sqrt{3} (z^6)^2 z^8 + 
\sqrt{3} (z^7)^2 z^8 - 6 z^3 (z^8)^2){\rm d}z^3  \nonumber \\
& + & (2(z^1)^2 z^4 + 2 (z^2)^2 z^4 + 2 (z^3)^2 z^4 - (z^4)^3 - z^4 (z^5)^2
- z^4 (z^6)^2 - z^4 (z^7)^2  \nonumber \\
&& - 2\sqrt{3} z^3 z^4 z^8 - 2\sqrt{3} z^1 z^6 z^8 + 2\sqrt{3} z^2 z^7 z^8)%
{\rm d}z^4  \label{SU3-2} \\
& + & (2(z^1)^2 z^5 + 2 (z^2)^2 z^5 + 2 (z^3)^2 z^5 - (z^4)^2 z^5 - (z^5)^3
- z^5 (z^6)^2 - z^5 (z^7)^2  \nonumber \\
&& - 2\sqrt{3} z^3 z^5 z^8 - 2\sqrt{3} z^2 z^6 z^8 - 2 \sqrt{3} z^1 z^7 z^8)%
{\rm d}z^5  \nonumber \\
& + & (2(z^1)^2 z^6 + 2(z^2)^2 z^6 + 2(z^3)^2 z^6 - (z^4)^2 z^6 - (z^5)^2
z^6 - (z^6)^3 - z^6 (z^7)^2  \nonumber \\
&& - 2\sqrt{3} z^1 z^4 z^8 - 2\sqrt{3} z^2 z^5 z^8 + 2\sqrt{3} z^3 z^6 z^8)%
{\rm d}z^6  \nonumber \\
& + & (2(z^1)^2 z^7 + 2 (z^2)^2 z^7 + 2 (z^3)^2 z^7 - (z^4)^2 z^7 - (z^5)^2
z^7 - (z^6)^2 z^7 - (z^7)^3  \nonumber \\
&& + 2\sqrt{3} z^8(z^2 z^4 - z^1 z^5 + z^3 z^7){\rm d}z^7] = 0 \, , 
\nonumber
\end{eqnarray}
where

\begin{eqnarray}
\Delta & = & 2 (z^1)^2 z^7 + 2 (z^2)^2 z^7 + 2 (z^3)^2 z^7 - (z^4)^2 z^7 -
(z^5)^2 z^7 - (z^6)^2 z^7 -(z^7)^3  \nonumber \\
&& +2\sqrt{3} (z^2 z^4 z^8 - z^1 z^5 z^8 + z^3 z^7 z^8) \, .  \label{SU3-3}
\end{eqnarray}
Neither equation (\ref{SU3-1}) nor (\ref{SU3-2}) can be integrated by
itself. So to obtain two independent Casimir invariants $C_1$ and $C_2,$ we
must find exact linear combinations of the equations. For example,

\begin{eqnarray}
z^8 \gamma ^{(1)} + z^7 \gamma ^{(2)} & = & z^1{\rm d}z^1 + z^2{\rm d}z^2 +
z^3{\rm d}z^3 + z^4{\rm d}z^4  \nonumber \\
&& + z^5{\rm d}z^5 + z^6{\rm d}z^6 + z^7{\rm d}z^7 +z^8{\rm d}z^8 = 0 \, .
\label{SU3-4}
\end{eqnarray}
This equation is exact, and has the solution

\begin{equation}  \label{SU3-5}
C_1 = (z^1)^2 + (z^2)^2 + (z^3)^2 + (z^4)^2 + (z^5)^2 + (z^6)^2 + (z^7)^2 +
(z^8)^2 \, .
\end{equation}
Another independent solution can be found by taking the linear combination

\begin{eqnarray}
a \gamma ^{(1)} + b \gamma ^{(2)} & = & (-18 z^4 z^6 - 18 z^5 z^7 - 12 \sqrt{%
3} z^1 z^8){\rm d}z^1  \nonumber \\
&+&(-18 z^5 z^6 + 18 z^4 z^7 - 12 \sqrt{3} z^2 z^8){\rm d}z^2  \nonumber \\
&+&(-9 (z^4)^2 - 9 (z^5)^2 + 9 (z^6)^2 + 9 (z^7)^2 -12 \sqrt{3} z^3 z^8){\rm %
d}z^3  \nonumber \\
&+&(-18 z^3 z^4 - 18 z^1 z^6 + 18 z^2 z^7 + 6 \sqrt{3} z^4 z^8){\rm d}z^4 
\nonumber \\
&+&(-18 z^3 z^5 - 18 z^2 z^6 - 18 z^1 z^7 + 6 \sqrt{3} z^5 z^8){\rm d}z^5
\label{SU3-6} \\
&+&(-18 z^1 z^4 - 18 z^2 z^5 + 18 z^3 z^6 + 6 \sqrt{3} z^6 z^8){\rm d}z^6 
\nonumber \\
&+&(18 z^2 z^4 - 18 z^1 z^5 + 18 z^3 z^7 + 6 \sqrt{3} z^7 z^8){\rm d}z^7 
\nonumber \\
&+&(-6 \sqrt{3} (z^1)^2 - 6 \sqrt{3} (z^2)^2 - 6 \sqrt{3} (z^3)^2 + 3 \sqrt{3%
} (z^4)^2  \nonumber \\
&& + 3 \sqrt{3} (z^5)^2 + 3 \sqrt{3} (z^6)^2 + 3 \sqrt{3} (z^7)^2 + 6 \sqrt{3%
} (z^8)^2){\rm d}z^8 =0 \, ,  \nonumber
\end{eqnarray}
where

\begin{eqnarray}
a & = & -3 \sqrt{3}[-2 (z^1)^2 - 2 (z^2)^2 - 2 (z^3)^2 + (z^4)^2 + (z^5)^2 +
(z^6)^2  \nonumber \\
&& + (z^7)^2 + 2 (z^8)^2],  \label{SU3-7} \\
b & = & 6[3 z^2 z^4 - 3 z^1 z^5 + 3 z^3 z^7 + \sqrt{3} z^7 z^8] \, .
\label{SU3-8}
\end{eqnarray}

Integrating equation (\ref{SU3-6}) yields the solution

\begin{eqnarray}
C_2 & = & -18 z^1 z^4 z^6 -18 z^1 z^5 z^7 - 6 \sqrt{3} (z^1)^2 z^8 +18 z^2
z^4 z^7 -18 z^2 z^5 z^6  \nonumber \\
&& - 6\sqrt{3} (z^2)^2 z^8 -9 z^3 (z^4)^2 - 9 z^3 (z^5)^2 + 9 z^3 (z^6)^2 +
9 z^3 (z^7)^2  \label{SU3-9} \\
&& - 6 \sqrt{3} (z^3)^2 z^8 + 3 \sqrt{3} (z^4)^2 z^8 + 3 \sqrt{3} (z^5)^2
z^8 + 3 \sqrt{3} (z^6)^2 z^8  \nonumber \\
&& + 3 \sqrt{3} (z^7)^2 z^8 + 2 \sqrt{3} (z^8)^3 \, .  \nonumber
\end{eqnarray}

We should note here that for Lie-Poisson brackets based on semi-simple Lie
algebras (as in this example), a formula exists for the Casimir invariants.
It follows immediately from results in \cite{gror} that if the basis
elements of the algebra are given as matrices $X^{(i)}$ in any
representation of the Lie algebra, then $C = Tr(X^{(i)} \cdots X^{(j)}) z^i
\cdots z^j$ is a Casimir invariant (here, $z^i$ is the coordinate dual to
the basis vector $X^{(i)}$). Using the Gell-Mann matrices in this formula
yields Casimir invariants proportional to (\ref{SU3-5}) and (\ref{SU3-9}).

As complicated as this last example might seem, the method used is certainly
more economical than solving the system of coupled PDEs mentioned in Section 
\ref{S-FINITE}, which is the only systematic alternative known for the
computation of Casimir invariants. And while for a small number of PDEs,
solutions can often be guessed, this becomes difficult for large systems.
The reader is referred to \cite{byv2} for additional examples.

\section{Casimir invariants of infinite-dimensional brackets}

\label{S-INFINITE}


For infinite-dimensional examples, we first consider Poisson brackets that
occur in the Hamiltonian formulation of the Vlasov-Poisson system and the 2D
Euler equation \cite{olv3,maip}. Then, we consider a modification of our
first example that might arise in other types of kinetic theories, such as
that which describes a spin gas.

All of the examples for this section arise in systems that can be described
by a single field variable $f.$ This field is taken to be a function on an $n
$-dimensional space D, with coordinates $z^i$, and of time. In these
examples, the space D is endowed with a Poisson bracket.

The brackets in these examples have the following general form: 
\begin{equation}  \label{INFBRAC}
\{ F,G \} [f] = \int _{{\rm D}} f \left[ {\frac{\delta F }{\delta f}}, {%
\frac{\delta G }{\delta f}} \right] {\rm d}^nz \, .
\end{equation}
In this expression, The variables $F, G$ represent functionals of the field $%
f$, and $\delta F / \delta f$ denotes the functional derivative of $F$ with
respect to $f$. The square brackets in the integrand denote the
(finite-dimensional) Poisson bracket defined on the space D. This will be
referred to as the inner bracket. This bracket is an infinite-dimensional
generalization of the Lie-Poisson bracket, first considered by Lie \cite{lie}%
. (The bracket is sometimes also known as the Kostant-Kirillov bracket \cite
{koki}.)

Furthermore, the brackets we consider can be rewritten in the form 
\begin{equation}  \label{INFCOSYM}
\{ F,G \} = -\int _{{\rm D}} {\frac{\delta F }{\delta f}} \left[ f, {\frac{%
\delta G }{\delta f}} \right] {\rm d}^n z \, .
\end{equation}
This is not true for all brackets of the form (\ref{INFBRAC}). Whether it
holds or not depends on the nature of the inner bracket, as will be seen
below.

Now, from equation (\ref{INFCOSYM}), we can read off the cosymplectic
operator as ${\rm J} = -[f, \cdot ] $. So the condition for finding the
components $\gamma $ of the null covector $\Gamma \colon = \gamma \delta f$
of J is simply

\begin{equation}  \label{INFNULL}
[f, \gamma ] = 0.
\end{equation}
To solve this completely for $\gamma $ we need to know the details of the
inner bracket. These will be provided for two cases below.


If the inner bracket in (\ref{INFBRAC}) is canonical, that is, takes the
form (\ref{CANBRAC}), an integration by parts (assuming that the field
vanishes on the boundary of D) shows that the bracket obeys equation (\ref
{INFCOSYM}). Hence, in this case, we can use equation (\ref{INFNULL}) to
determine its Casimir invariants.

By antisymmetry of the bracket, $\gamma =f$ solves equation (\ref{INFNULL}).
Hence $\gamma = k(f)$ is also a solution, for an arbitrary function $k.$
Since the variation $k(f) \delta f$ is an exact variation (the formal
analogue in infinite dimensions of an exact differential), the Pfaffian
system, $k(f) \delta f = 0,$ is easily integrated. This yields the
expression for the Casimir invariant ${\cal C}$:

\begin{equation}  \label{INFCAS.1}
{\cal C}[f] = \int _D K(f) {\rm d}^n z \, ,
\end{equation}
where $K(f)$ is the primitive with respect to $f$ of $k(f).$ (Note: we use
the caligraphic font to distinguish Casimir invariants of
infinite-dimensional brackets from those of finte dimensional brackets.)


As another, slightly more complicated example, we consider a bracket of the
form (\ref{INFBRAC}) with an inner bracket of Lie-Poisson type. We suggest
that brackets of this form are applicable to the Hamiltonian formulations of
nontraditional kinetic theories in analogy to their use in the
Vlasov-Poisson system mentioned above. Notice that the inner bracket for the
Vlasov-Poisson system is that appropriate to the phase space of a point
particle. If, however, we wished to describe a ``gas'' composed of freely
spinning (fixed) rigid bodies instead of point particles, we expect that the
inner bracket for this kinetic theory would be the so(3) bracket mentioned
in Section \ref{S-FINITE}. Another, more exotic, possibility would be a
kinetic theory that describes a gas of Kida vortices. The dynamics of a Kida
vortex takes place on the Lie algebra so(2,1) (see \cite{mmof}), and is
governed by the corresponding Lie-Poisson bracket.

\begin{equation}  \label{ALGBRAC}
[f, g] (z) = c_i^{jk} z^i {\frac{\partial f }{\partial z^j}} {\frac{\partial
g }{\partial z^k}} \, ,
\end{equation}
where $c_i^{jk}$ are the structure constants of the Lie coalgebra.

Integration by parts (assuming $f$ is constant on the boundary) shows that
the bracket (\ref{INFBRAC}) with inner bracket of form (\ref{ALGBRAC})
satisfies equation (\ref{INFCOSYM}) if and only if the structure constants
obey the condition $c_i^{ik} = 0$ for all $k.$ This happens to be satisfied
by the structure constants of any semi-simple Lie algebra. (Incidentally,
any Hamiltonian flow generated by a Lie-Poisson bracket satisfying this
condition obeys the naive Liouville's theorem in noncanonical coordinates 
\cite{mrmp}.)

Now, to calculate the Casimir invariants for these examples, we again must
solve equation (\ref{INFNULL}) for $\gamma $. As in the above example, one
solution is given by $\gamma = f.$ However, the inner bracket in this
example {\it can} be degenerate, and thus have Casimir invariants of its
own. Suppose there are $m$ such independent functions and denote them by $%
C^{(i)}.$ Then a solution of (\ref{INFNULL}) is $\gamma =
k(C^{(1)},\ldots,C^{(m)},f),$ where $k$ is an arbitrary function. From this
we see that

\begin{equation}  \label{GENINFNULL}
\Gamma = k(C^{(1)},\ldots,C^{(m)},f) \delta f
\end{equation}
is a null covector of J. Not only that, it is an exact variation, making $%
\Gamma = 0$ easy to integrate. Indeed, we find as a Casimir invariant

\begin{equation}  \label{INFCAS}
{\cal C}[f] = \int _{D} K(C^{(1)},...,C^{(m)},f) {\rm d}^nz
\end{equation}
where $k(f) =\partial K(f)/\partial f.$

Applying this result to the proposed spin gas bracket, we use the result
from Section \ref{S-FINITE}, and obtain the Casimir invariant

\begin{equation}  \label{INFSO3CAS}
{\cal C}[f] = \int _{D} K((z^1)^2 + (z^2)^2 + (z^3)^2,f) {\rm d}^3z \, .
\end{equation}
We could also envision a kinetic theory describing particles evolving on the
Lie algebra su(3). Its Casimir invariant would be: 
\begin{equation}  \label{INFSU3CAS}
{\cal C}[f] = \int _{D} K(C_1, C_2,f) {\rm d}^8z \, ,
\end{equation}
where $C_1, C_2$ are given respectively by equations (\ref{SU3-5}, \ref
{SU3-9}). In both (\ref{INFSO3CAS}) and (\ref{INFSU3CAS}), $K$ is an
arbitrary function.

Note the distinct origins of the two types of functional dependence $K$ in
these invariants. Its dependence on $f$ arises from the structure of the
infinite-dimensional bracket alone, and so is common to any kinetic theory
to which bracket (\ref{INFBRAC}) is applicable: it might be called the
``macroscopic'' part of the Casimir invariant. On the other hand, the
dependence of $K $ on the $C^{(i)}$ arises from the structure of the
finite-dimensional inner bracket, and so reflects the nature of the
``microscopic'' theory.

\section{Conclusion}

\label{S-CONCL}

We have presented a method for computing Casimir invariants applicable to
both finite and infinite-dimensional Poisson brackets. We believe that
presenting this method geometrically has two benefits: it enhances pictorial
intuition of the behavior of finite-dimensional Hamiltonian systems, and it
gives insight into how infinite-dimensional Hamiltonian systems can be
treated analogously to finite-dimensional ones.

We have also computed some new examples of Casimir invariants. Most
interesting, perhaps, are those that are the results of Poisson structures
at finite and infinite-dimensional levels simultaneously. In these, we have
pointed out what parts of the invariants correspond to which Poisson
structure.

\begin{flushleft}
{\bf Acknowledgements}
\end{flushleft}

BH acknowledges Comunidad Aut\'{o}noma de Madrid for financial support, as
well as the Institute of Fusion Studies of the University of Texas at Austin
for their kind hospitality. TWY would like to thank Jean-Luc Thiffeault for
many valuable conversations. PJM acknowledges the hospitality of the GFD
Summer School held at the Woods Hole Oceanographic Institute. TWY and PJM
are supported by the Department of Energy under Contract No.
DE-FG03-96ER-54346.

\pagebreak

\end{document}